\documentclass[twocolumn,aps,superscriptaddress,pre]{revtex4}

\usepackage{amsmath,amssymb,graphicx}
\usepackage{algorithmic}
\usepackage{enumerate}
\usepackage{color}

\definecolor{yblue}{rgb}{0.06, 0.3, 0.57}
\usepackage[pdftex]{hyperref}
\hypersetup{colorlinks=true,linkcolor=blue,citecolor=blue,urlcolor=blue}

\newcommand{\ds}{d_{\rm s}/2}

\begin{document}

\title{Chaotic temperature and bond dependence of four-dimensional Gaussian spin glasses with partial thermal boundary conditions}

\author{Wenlong Wang}
\email{wenlongcmp@gmail.com}
\affiliation{Department of Physics, Royal Institute of Technology, Stockholm, SE-106 91, Sweden}

\author{Mats Wallin}
\affiliation{Department of Physics, Royal Institute of Technology, Stockholm, SE-106 91, Sweden}

\author{Jack Lidmar}
\affiliation{Department of Physics, Royal Institute of Technology, Stockholm, SE-106 91, Sweden}

\begin{abstract}

Spin glasses have competing interactions and complex energy
landscapes that are highly-susceptible to perturbations, such as the temperature or the bonds. %These chaotic effects strongly affect numerical simulations and experiments. %, as such, gaining a deeper understanding of chaos in spin glasses is of significant importance. 
The thermal boundary condition technique is an effective and visual approach for characterizing chaos, and has been successfully applied to three dimensions. In this paper, we tailor the technique to partial thermal boundary conditions, where thermal boundary condition is applied in a subset (3 out of 4 in this work) of the dimensions for better flexibility and efficiency for a broad range of disordered systems. We use this method to study both temperature chaos and bond chaos of the four-dimensional Edwards-Anderson model with Gaussian disorder to low temperatures. We compare the two forms of chaos, with chaos of three dimensions, and also the four-dimensional $\pm J$ model. We observe that the two forms of chaos are characterized by the same set of scaling exponents, bond chaos is much stronger than temperature chaos, and the exponents are also compatible with the $\pm J$ model. Finally, we discuss the effects of chaos on the number of pure states in the thermal boundary condition ensemble.
\end{abstract}

%\pacs{75.50.Lk, 75.40.Mg, 05.50.+q, 64.60.-i}
\maketitle

\section{Introduction}

Chaos is a fascinating and common phenomenon in glassy systems, which have rugged energy landscapes such as spin glasses. The spin orderings are reorganized at large scales when a parameter is tuned, such as the temperature or the bonds. These corresponding chaotic phenomena are therefore called temperature chaos \cite{mckay:82,kondor:89,parisi:84,fisher:86,bray:87,ritort:94,rizzo:03,rizzo:06,
sasaki:03,Hukushima:Chaos,Katzgraber:Chaos,thomas:11e,
fernandez:13,Cecile:Chaos,Wang:TC,Parisi:TC} and bond chaos \cite{Hukushima:Chaos,Katzgraber:Chaos,Cecile:Chaos,Wang:BC}, respectively.
While chaos is an equilibrium phenomena, it is also believed to be related to various non-equilibrium dynamics such as hysteresis, memory and rejuvenation effects \cite{fisher:91d,sales:02,silveira:04,Doussal:Chaos}.
Chaos is also of great relevance for numerical simulations and
analog optimization machines \cite{Zhu:Chaos,Hen:QA}, such as the D-Wave quantum annealers. For example, small temperature perturbations or problem misspecifications could lead to a solution of an entirely different Hamiltonian, especially when the number of spins is large.
Chaos is a source of the computational complexity of spin glasses \cite{fernandez:13,Wang:TC,Hen:QA,AB:chaos}, known to slow down extended-ensemble algorithms, which are the current state-of-the-art methods, including both parallel tempering and population annealing.
Therefore, chaos is closely related to both equilibrium and nonequilibrium properties of spin glasses, experimental optimizations, and numerical simulations.

It has been recognized that temperature chaos (TC) and bond chaos (BC) appear to follow the same scaling properties, and bond chaos is considerably stronger than temperature chaos \cite{krzakala:05,Hukushima:Chaos,Katzgraber:Chaos}. Both of these results can be simply explained within the framework of the droplet picture \cite{fisher:86,fisher:87,fisher:88,bray:86,mcmillan:84b} by scaling properties and assuming that temperature chaos is mainly entropy driven, whereas bond chaos is mainly energy driven \cite{Wang:BC}.
%\cite{fisher:86,fisher:87,fisher:88,bray:86,mcmillan:84b}

Most studies of chaos are based on some correlation functions \cite{neynifle:97,neynifle:98,krzakala:05,Hukushima:Chaos,Katzgraber:Chaos}. Recently, a new technique called thermal boundary conditions (TBC) has been successfully applied to three-dimensional spin glasses \cite{Wang:TC,Wang:BC}. For thermal boundary conditions, the system can choose either periodic or antiperiodic boundary conditions in each spatial direction, according to the Boltzmann weights of the different boundary conditions. In D dimensions, the full TBC set has $2^{\rm{D}}$ different boundary conditions.
Chaos manifests itself as the instabilities of the relative weights of different boundary conditions (in thermal equilibrium) when the temperature or the bonds are tuned.

The TBC approach has certain advantages. Firstly, the strength of chaos is directly quantified using number of boundary condition crossings (exchange of their weights). Therefore, there is no reference state such as a reference temperature as in correlation functions. This allows a direct and detailed characterization of chaos such as the temperature dependence of the strength of temperature chaos. Chaotic events are also more frequently observed with the enlarged phase space, with some chaotic instances exhibiting several crossings in a typical parameter range (such as a temperature range for temperature chaos) even for a relatively small system size accessible to current simulations.

%and what is the relative strength of temperature chaos and bond chaos at a fixed temperature?
%Finally, the only observables needed are energies and relative weights of the set of boundary conditions (or equivalently free energies which can be efficiently measured using the free-energy perturbation method), which converges much faster than many other variables, such as the spin overlap order parameter.

Despite of these successes and extensive research of chaos in three dimensions, there are far less work in four dimensions \cite{neynifle:98,Hukushima:Chaos,ritort:94} and the majority of these works focused on the $\pm J$ model \cite{ritort:94,Hukushima:Chaos}. To the best of our knowledge, we have only found one such pioneering numerical study on the Gaussian disorder in four dimensions operating at a relatively high temperature using correlation functions \cite{neynifle:98}. This is most likely due to earlier computational limitations, considering that Gaussian disorder is \textit{much} harder to equilibrate than the $\pm J$ disorder. In this paper, we fill in this gap and study the numerically intensive four-dimensional Gaussian spin glasses to low temperatures ($T_C/3$ for temperature chaos and $T_C/2$ for bond chaos) using the massively-parallel algorithm population annealing. This not only improves statistical errors for a better comparison of temperature chaos and bond chaos in 4D, but more importantly also allows us to compare with the 3D counterpart, and the 4D $\pm J$ model. Secondly, we also tailor the TBC technique to apply more flexibly and efficiently to the 4D model (and many others, e.g., the one-dimensional chains with long-range interactions). Our work is done using \textit{partial} thermal boundary conditions which is described as follows.

%Because of these features, in this work we study \textit{efficient} extensions of the technique to more systems. 
The motivation for the partial thermal boundary condition is from the following question: Is the total number of boundary conditions essential to the TBC technique? For example, is it necessary to keep all $16$ boundary conditions in 4D, which is a rather expensive setup? Much computational efforts would be saved if we could reduce this number. On the other hand, for a one-dimensional spin chain with long-range interactions, one would like to use more boundary conditions rather than two to collect good statistics. In this work, we propose a simple idea to tailor the number of boundary conditions. More precisely, we introduce the \textit{partial thermal boundary conditions} in four dimensions, to turn on thermal boundary conditions in only a subset of the dimensions. As mentioned, to collect good statistics, the number of boundary conditions should also not be too small. Therefore, we choose to keep $8$ boundary conditions as in 3D, i.e., thermal boundary condition is turned on in three directions and periodic boundary condition is always applied in the fourth direction. There could be a potential possibility that changing the number of boundary conditions may affect the scaling exponent of the number of crossings. Fortunately, our results suggest this is not the case and the method is valid, as shown in Sec.~\ref{results}.

The paper is organized as follows. We first present the model,
simulation methods and scaling properties of temperature chaos and bond chaos in Sec.~\ref{mm}, followed by numerical results in Sec.~\ref{results}. Concluding remarks are stated in Sec.~\ref{cc}.

%We also propose $\zeta=ds/2-\theta$ by looking at the zero temperature exponents $ds$ and $\theta$. The exponent $\theta$ has been mapped out up to $d=6$ for a few years using percolation method, but it was only until recently that $ds$ was measured up to $d=8$ using the strong disorder renormalization group method. One can actually see that this is not really true. i.e. $\zeta$ is not a constant of the dimensionality. Actually if $\theta=1$ and $ds=d$ for $d>6$, one can expect $\zeta(d)=d/2-0.5$ for $d>6$. We present numerical values of $\zeta$ is presented up to $d=8$.

\section{Models and numerical setup}
\label{mm}

In this Section, we present the four-dimensional Edwards-Anderson model, observables and simulation details. The scaling properties for characterizing the chaos phenomena are also summarized for completeness.

\subsection{Models, methods and observables}

The Edwards-Anderson (EA) Ising spin glass \cite{EA} is
represented by the following Hamiltonian:
\begin{equation}
H = -\sum\limits_{\langle ij \rangle} J_{ij} S_i S_j,
\end{equation}
where $S_i \in \{\pm 1\}$ are Ising spins. The sum $\langle ij \rangle$
is over the nearest neighbours in a four-dimensional simple cubic lattice of linear system size $L$ and number of spins $N=L^4$.
The couplings $J_{ij}$ between spins $S_i$ and $S_j$ are chosen independently from the standard Gaussian distribution with mean zero and variance one. We refer to each disorder realization as an ``instance''. We apply partial thermal boundary conditions (PTBC) to each instance, i.e., each instance has freedom to choose either periodic boundary conditions or antiperiodic boundary conditions in three directions according to the Boltzmann weights. In the fourth direction, periodic boundary condition is always applied. There are therefore a total of eight boundary conditions in our PTBC ensemble. More precisely, the weight $p_i$ of a boundary condition $i$ is related to its free energy $F_i$ as:
\begin{eqnarray}
p_i = \frac{\exp(-\beta F_i)}{\sum_i \exp(-\beta F_i)}.
\end{eqnarray}

\begin{figure}[htb]
\begin{center}
\includegraphics[width=\columnwidth]{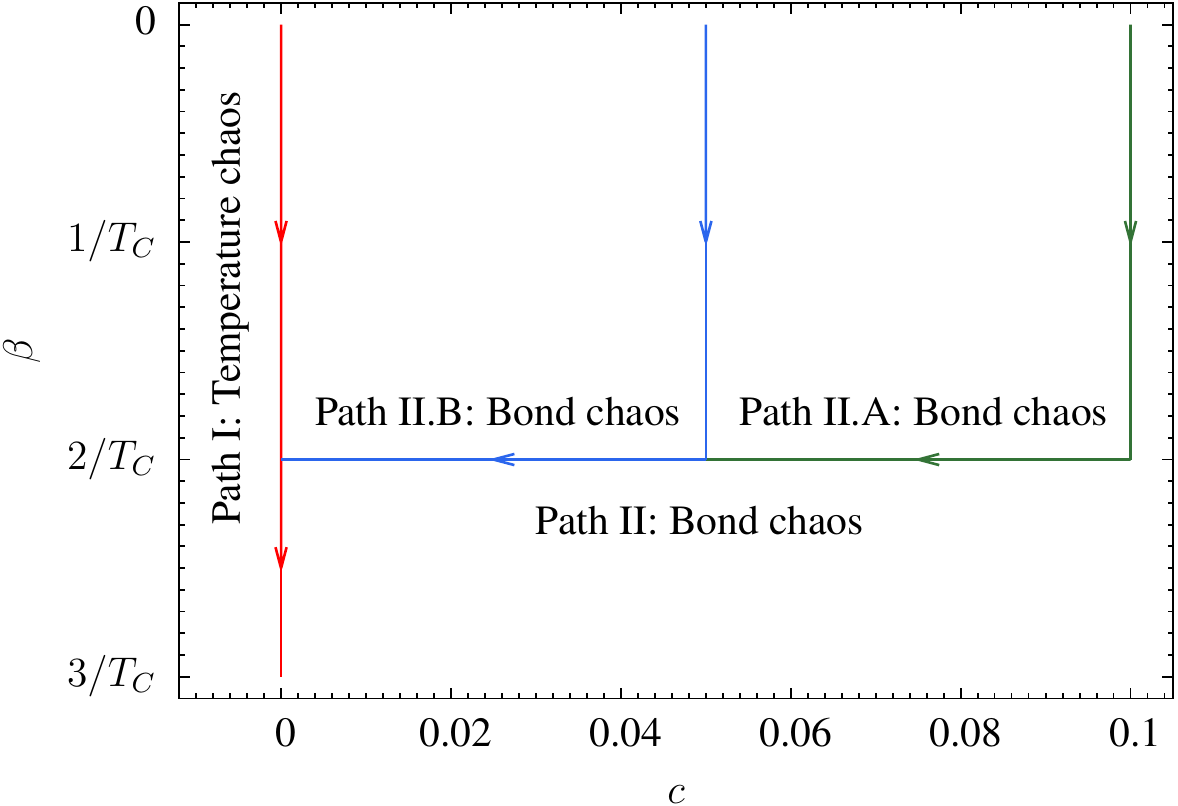}
\caption{
Schematic simulation paths for temperature chaos and bond chaos. In all cases, an annealing from $\beta = 0$ to $T=T_C/2$ is performed. For temperature chaos, the path goes straight down in temperature reaching $T=T_C/3$. For bond chaos, the path turns horizontally into the $c$ direction at a constant $T=T_C/2$. When equilibration criteria are not met for an instance, we rerun it with a larger population size or more sweeps. However, if an instance is too chaotic to equilibrate in the $c$-path, then spending more computational work becomes impractical. The path is then split into two (or more, two are shown here) paths, which are later combined to form the full path for data analysis.
}
\label{path}
\end{center}
\end{figure}

The model has a spin-glass phase transition at $T_C \approx 1.8$ \cite{parisi:96,neynifle:98}. For later references, we mention here that the 3D Gaussian model has $T_C \approx 1$ \cite{katzgraber:06} and the 4D $\pm J$ model has instead $T_C \approx 2$ \cite{marinari:99}. To study temperature chaos, a single instance $\mathcal{J}$ is cooled from the infinite temperature $\beta=0$ to a low temperature deep in the spin-glass phase $T_C/3$. Scaling properties are studied in the temperature range $T \in [T_C/3, 2T_C/3]$. To study bond chaos, we first choose an independent random perturbation instance $\mathcal{J}'$ for each instance $\mathcal{J}$. We then tune the bonds using a small parameter $c$ at a fixed temperature $T_C/2$ following an annealing also from $\beta=0$ as:
\begin{equation}
J=\dfrac{{\mathcal J}+c{\mathcal J}'}{\sqrt{1+c^2}},
\end{equation}
where $c \in[0,0.1]$. The normalization factor is to preserve the standard Gaussian distribution for any $c$ \cite{neynifle:97,neynifle:98,krzakala:05,Katzgraber:Chaos,Wang:BC}. Note that the possibility to change the Gaussian bonds continuously over a range is a convenient advantage against discrete bonds such as the $\pm J$ model \cite{Hukushima:Chaos}. In our simulations, we start from $c=0.1$ and then reduce $c$ to 0, and the final instance becomes $\mathcal{J}$. Note that the final $\mathcal{J}$ is chosen to be identical to the temperature chaos instance for benchmarking purposes as equilibrium properties should not depend on how the system is prepared. The simulations can be clearly visualized by looking at the simulation trajectories in the parameter space $(\beta,c)$ in Fig.~\ref{path}.

Our simulation is carried out using population annealing Monte Carlo \cite{Hukushima:PA,Zhou:PA,Machta:PA,Wang:PA,Weigel:PA}. For each instance, we initialize $R$ random replicas each with a random configuration and a random boundary condition at $\beta=0$. Define $\mathcal{H}=\beta H$ as the reduced Hamiltonian. When we change the simulation parameters as in Fig.~\ref{path}, or the reduced Hamiltonian from $\mathcal{H}$ to $\mathcal{H}'$, a replica $i$ is copied with the expectation number $n_i=\exp[-(\mathcal{H}_i'-\mathcal{H}_i)]/Q$. Here, $Q=(1/R)\sum_i \exp[-(\mathcal{H}_i'-\mathcal{H}_i)]$ is a normalization factor to keep the population size approximately the same as $R$. In our simulation, the number of copies is randomly chosen as either the floor or the ceiling of $n_i$ with proper probabilities to minimize fluctuations. This reweighting step is called resampling in population annealing, and note that some replicas would be duplicated while others may get eliminated from the population. The purpose of the resampling is to try to maintain the population in equilibrium when simulation parameters are changed. After this resampling step, $N_S$ sweeps using the Metropolis algorithm is applied to each replica. The annealing process continues with the cyclic resampling and Monte Carlo sweeps until the final targeted parameters are reached. More details on simulation methods can be found in the three-dimensional work~\cite{Wang:TC,Wang:BC}. %Other simulation methods, as well as simulating the TBC ensemble using parallel tempering, can also be found in Ref.~\cite{Wang:PTTBC}. 
The simulation parameters are summarized in Table.~\ref{para}.

%Key ingredients of the paper:
%(1) Measure the three exponents for both TC and BC for the 4D EA model, and check if they are the same as well if $\zeta=ds/2-\theta$ is satisfied.
%(2) We validate the idea of partial thermal boundary conditions.
%(3) We map out $\zeta$ as a function of dimensionality and give an expected relation for $d>6$.
%(4) Compare with 3D such as does chaos gets stronger in 4D?

\begin{table}
\caption{
Simulation parameters of chaos for the four-dimensional spin glasses using population annealing. $\mathcal{BC}$ is the boundary conditions, $\mathcal{TB}$ are either for temperature chaos (TC) or bond chaos (BC), $L$ is the
linear system size, $R$ is the number of replicas or population size, $T_{\rm{min}}$ is the lowest temperature simulated, $N_T$ is the number of temperature steps (evenly spaced in $\beta$) and $N_c$ is the number of disorder steps (evenly spaced in $c$) in the annealing schedule, and $M$ is the number of instances studied. We apply $N_S=10$ sweeps to each replica after each annealing step.
\label{para}
}
\begin{tabular*}{\columnwidth}{@{\extracolsep{\fill}} c c c c c c c c}
\hline
\hline
$\mathcal{BC}$ &$\mathcal{TB}$ &$L$  &$R$  &$T_{\rm{min}}$ & $N_T$ & $N_c$ & $M$ \\
\hline
PTBC &TC &$4$  &$2\times10^5$ &$0.6$  &$101$ &- &$2000$ \\
PTBC &TC &$5$  &$6\times10^5$ &$0.6$  &$101$ &- &$2000$ \\
PTBC &TC &$6$  &$8\times10^5$ &$0.6$  &$201$ &- &$2000$ \\
PTBC &TC &$7$  &$2\times10^6$ &$0.6$  &$301$ &- &$2000$ \\
PTBC &BC &$4$  &$2\times10^5$ &$0.9$  &$101$ &$51$ &$2000$ \\
PTBC &BC &$5$  &$6\times10^5$ &$0.9$  &$101$ &$51$ &$2000$ \\
PTBC &BC &$6$  &$8\times10^5$ &$0.9$  &$201$ &$51$ &$2000$ \\
PTBC &BC &$7$  &$2\times10^6$ &$0.9$  &$301$ &$101$ &$2000$ \\
%PBC &- &$4$  &$5\times10^4$ &$0.5$  &$101$ &- &$2000$ \\
%PBC &- &$5$  &$5\times10^4$ &$0.5$  &$101$ &- &$2000$ \\
%PBC &- &$6$  &$5\times10^4$ &$0.5$  &$101$ &- &$2000$ \\
%PBC &- &$7$  &$5\times10^4$ &$0.5$  &$101$ &- &$2000$ \\
%APBC &- &$4$  &$5\times10^4$ &$0.5$  &$101$ &- &$2000$ \\
%APBC &- &$5$  &$5\times10^4$ &$0.5$  &$101$ &- &$2000$ \\
%APBC &- &$6$  &$5\times10^4$ &$0.5$  &$101$ &- &$2000$ \\
%APBC &- &$7$  &$5\times10^4$ &$0.5$  &$101$ &- &$2000$ \\
\hline
\hline
\end{tabular*}
\end{table}

Our equilibration criteria are based on a combination of family entropy, and matching of boundary condition weights when two simulation paths meet in the $(\beta,c)$ parameter space \cite{Wang:PA,Wang:BC}. Note that we test equilibration for each individual instance rather than the disorder average of all instances. Copying replicas reduces the diversity of the population, and family entropy quantifies this property. In the initial population, each replica is given a family name $1, 2, 3, ..., R$. A family name is copied together with a replica when doing resamplings and remains the same under Monte Carlo updates. At each stage of the simulation, we collect the fraction of each family name in the population $\{f_i\}$ and the family entropy $S_f$ is then defined using the regular Gibbs entropy $S_f = -\sum_i f_i \ln (f_i)$ \cite{Wang:TBC,Wang:PA}. The family entropy usually decreases as simulation proceeds, and it is sufficient to control the final family entropy of each instance \cite{Wang:TBC,Wang:PA}. The final family entropy depends on the energy landscape of an instance and the simulation details such as the population size and the number of sweeps. The larger $S_f$, the better the equilibration for a simulation. We require each simulation to satisfy $S_f \geq \ln(100)$. Whenever two simulation paths meet in the parameter space, we require also that the two simulation paths should give the same boundary condition weights 
$\max\{|p_i-p_i'|\} \leq 0.05$, where $\{p_i\}$ and $\{p_i'\}$ are the weights of each boundary conditions from the two paths, respectively. When either criterion is not fulfilled for an instance by using the parameters in Table.~\ref{para}, we rerun it either by increasing the population size, doing more sweeps, or breaking the $c$ path into several segments, as shown in Fig.~\ref{path}.

Finally, for each instance, we record the energy $\{E_i\}$ and weights $\{p_i\}$ of each boundary condition along the simulation paths. The energy is computed by averaging over the replicas and the weights are estimated by counting the fraction of replicas of each boundary condition. All other observables used for studying chaos in this work are derived from only these two observables, reflecting the simplicity of the method. Other observables not directly related to chaos such as the free energy and the order parameter or the overlap distribution function will be defined when used for clarity. We summarize the scaling properties of chaos in the next section.

\subsection{Scaling analysis}
\label{sa}
In this section, we summarize the scaling relations used in this work in the framework of the droplet picture. Flipping boundary conditions would create a relative domain wall between two boundary conditions. There are two scaling exponents in the droplet picture for such domain walls: the domain-wall free energy exponent $\theta$ and the domain-wall fractal dimension $d_s \in [d-1,d]$. Let $\Delta F$ be the free energy cost of inserting a domain wall and $\Sigma$ is the size or number of spins of the domain wall, then
\begin{eqnarray}
\label{DF}
\Delta F &\sim & L^{\theta}, \\
\Sigma &\sim & L^{d_s}.
\label{ds}
\end{eqnarray}

Naturally $\Delta F=0$ at a boundary condition crossing, but both $\Delta E$ and $T\Delta S$ are nontrivial like in a first-order phase transition and they scale as:
%In the vicinity of a boundary condition crossing, naturally $\Delta F=0$, at the crossing both energy cost and entropy cost scales as
\begin{eqnarray}
\label{DE}
\Delta E &\sim & L^{d_s/2}, \\
\label{DS}
T \Delta S &\sim & L^{d_s/2}.
\end{eqnarray}
Here it is simply assumed the scales are related to the size of domain walls (Eq.~\ref{ds}) and the square roots come from the frustrations of domain walls. Doing a Taylor expansion in the vicinity of a crossing  for the generalized parameter $Q$ at $Q_0$ (either $Q=T$ for temperature chaos or $Q=c$ for bond chaos) gives:
\begin{eqnarray}
\Delta F (Q_0+\delta Q) &=& \Delta F (Q_0) + \dfrac{\partial \Delta F}{\partial Q} \delta Q. \\
&=& \dfrac{\partial (\Delta E - T\Delta S)}{\partial Q} \delta Q.
\end{eqnarray}
Suppose that $\Delta E$ dominates the response to bond changes and $T\Delta S$ dominates the response to temperature changes \cite{Wang:BC}, we obtain:
\begin{eqnarray}
L^{\theta} &\sim & L^{d_s/2} \delta Q, \\
\label{DQ}
\delta Q &\sim & 1/L^{\zeta}, \\
\delta Q &\sim & 1/L^{d_s/2 - \theta},
\end{eqnarray}
where $\zeta = d_s/2 -\theta$ is the chaos exponent. Note that this is a derived exponent that depends on $d_s$ and $\theta$. In this work, we measure these three exponents \textit{independently} for both forms of chaos and check this equality.
One direct consequence of Eq.~\ref{DQ} is that the number of dominant boundary condition crossings $N_C$ should scale as:
\begin{equation}
\label{NC}
N_C \sim L^{\zeta},
\end{equation}
where a dominate boundary condition crossing is a crossing of two boundary conditions that also have the maximum weights. See the red circles in Fig.~\ref{BCC} for examples.

The exponent $\theta$ can also be measured in the framework of thermal boundary conditions using the so-called sample stiffness scaling \cite{Wang:TBC,Wang:TC}. In this approach, free energy is not measured directly like energy, although this is also possible using the free energy perturbation method \cite{Hukushima:PA,Wang:PA}. Rather domain-wall free energy is conveniently estimated from the quantity sample stiffness. For an instance at a temperature $T$, it is defined as:
\begin{equation}
\label{eq:lam}
\lambda (T) = \log\frac{p_{\rm{max}}(T)}{1-p_{\rm{max}}(T)},
\end{equation}
where $p_{\rm{max}}=\max(\{p_i\})$ is the
maximum weights of all the boundary conditions.
Note that this is simply an estimator of the free-energy difference (times $-\beta$) between the dominant boundary condition and all other
boundary conditions combined. Since $p_{\rm{max}}$ can be very close to 1 for some instances, and a precise estimation of $\lambda$ for these instances would be difficult, one therefore usually works with a characteristic $\lambda_{\rm char}$ using a median, instead of the mean. The median is usually chosen from the tail of the distribution (large $\lambda$), but not too far into the tail where statistics are poor. In our work, we choose the 0.9 median and we have checked that our results are not sensitive to this particular choice. Naturally as Eq.~\ref{DF}, $\lambda_{\rm char}$ scales as:
\begin{equation}
\label{theta}
\lambda_{\rm char} \sim L^{\theta}.
\end{equation}

%Let $G_L(\lambda)$ be the
%cumulative distribution function for $\lambda$, then it was shown in
%Ref.~\cite{Wang:TBC} that the function $1-G_L(\lambda)$ is approximately
%exponential, which then allows for a scaling analysis. Define a
%characteristic $\lambda_{\rm char}(L)$ such that $1-G(\lambda) =
%e^{-\lambda/\lambda_{\rm char}}$ and $1-G(\lambda_{\rm char} \log b) =
%1/b$ for any $b$. The value $b$ should be chosen such that $\lambda_{\rm char}$ is obtained from the tail of the distribution but not so far into
%the tail where the statistics are poor. 
%In our work, we choose $b = 10$ and we have verified that the distribution functions for different linear system sizes $L$ collapse well onto the same curve after being scaled by $\lambda_{\rm char}(L)$. Note that $\lambda_{\rm{char}}$ is just the $1-1/b=0.9$ median of $\lambda$, upto a factor of $\log(b)=\log(10)$. Similar to the regular spin stiffness, it scales as the free energy.

%In the thermal dynamic limit, it is likely the two exponents are the same. In simulations for finite sizes, however, it was found $\theta_{\rm{SS}}$ is larger than $\theta$ \cite{Wang:TBC}. It was conjectured $\theta_{\rm{SS}}$ is closer to the thermal dynamic limit as the free energy scaling are more dominated by large free differences. In this work, we measure this exponent for completeness. For our study of chaos for our sizes, the spin stiffness exponent between \textit{two} boundary conditions is more relevant. In the next section, we present our numerical results on temperature chaos and bond chaos.

We summarize our methods for measuring the scaling exponents: We use sample stiffness scaling (Eq.~\ref{theta}) to measure $\theta$. At the boundary condition crossings $\Delta F = 0$, and we use $\Delta E$ (Eq.~\ref{DE}) to measure $\ds$. We use only crossings that are above a threshold for good accuracy. For temperature chaos, we use crossings above $p_c=0.05$. For bond chaos where there are more crossings, we use a slightly larger threshold $p_c=0.1$. Our results, however, are not sensitive to these thresholds. We use the number of dominant crossings $N_C$ (Eq.~\ref{NC}) to compute the exponent $\zeta$. Note that the different thresholds do not affect $N_C$, as no dominant crossings can occur below $p=0.125$ with 8 boundary conditions. In the next section, we present our results of temperature chaos and bond chaos, and the comparisons with the 3D model and the 4D $\pm$ J model.
%using our PBC and APBC data, where free energy is measured using the free energy perturbation method. We use the scaling of Eq.~\ref{DE} (or equivalently Eq.~\ref{DS}) at all registered crossings (above $p=0.05$). Finally, we use Eq.~\ref{NC} to compute $\zeta$ using dominate crossings. Note that to measure $d_s$, one can also use only dominate crossings, but we use all crossings to reduce statistical errors. It is, however, important to use dominate crossings, rather than all resisted crossings for $\zeta$ to reduce finite size effect, as crossings are not registered for $p$ smaller than the chosen threshold, where weights of a boundary condition may become very small to be measured accurately.

\section{Results}
\label{results}

\subsection{Scaling properties of chaos}
\begin{figure}[htb]
\begin{center}
\includegraphics[width=\columnwidth]{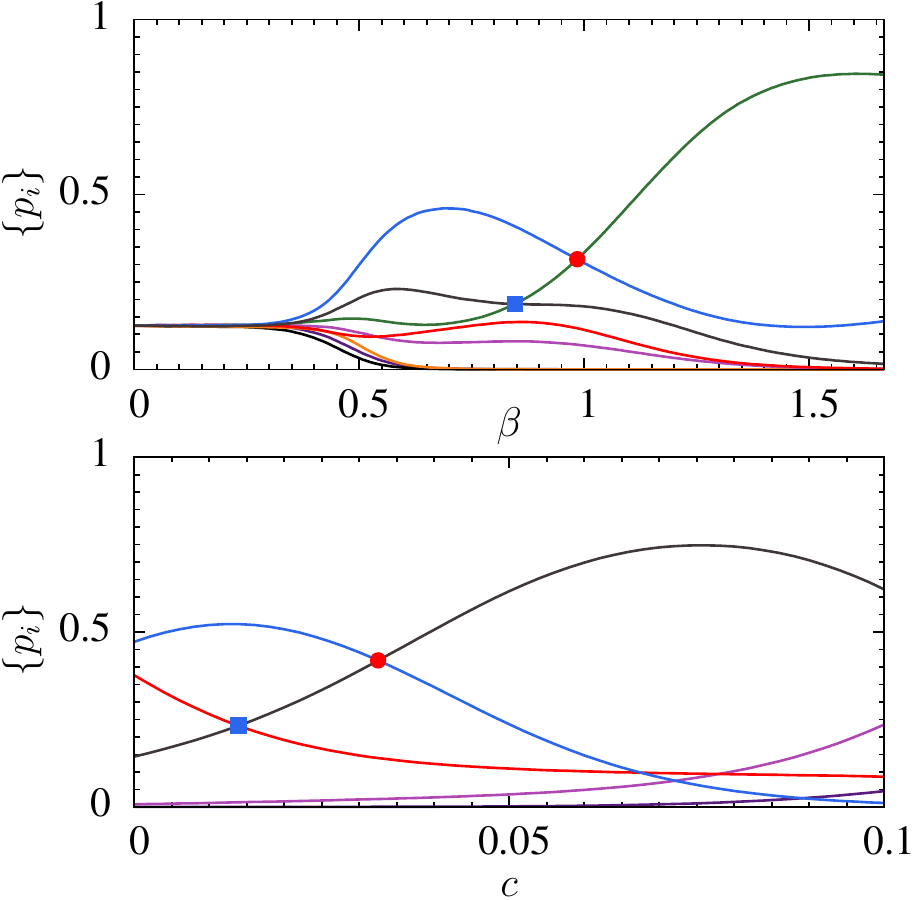}
\caption{
Two typical evolutions of the weights $\{p_i\}$ of each boundary condition of two moderately chaotic instances of system size $L=6$ for temperature chaos (upper panel) and bond chaos (lower panel), respectively. The red circles are examples of dominant crossings, and the blue squares are crossings but not dominant ones.
}
\label{BCC}
\end{center}
\end{figure}

\begin{figure}[htb]
\begin{center}
\includegraphics[width=\columnwidth]{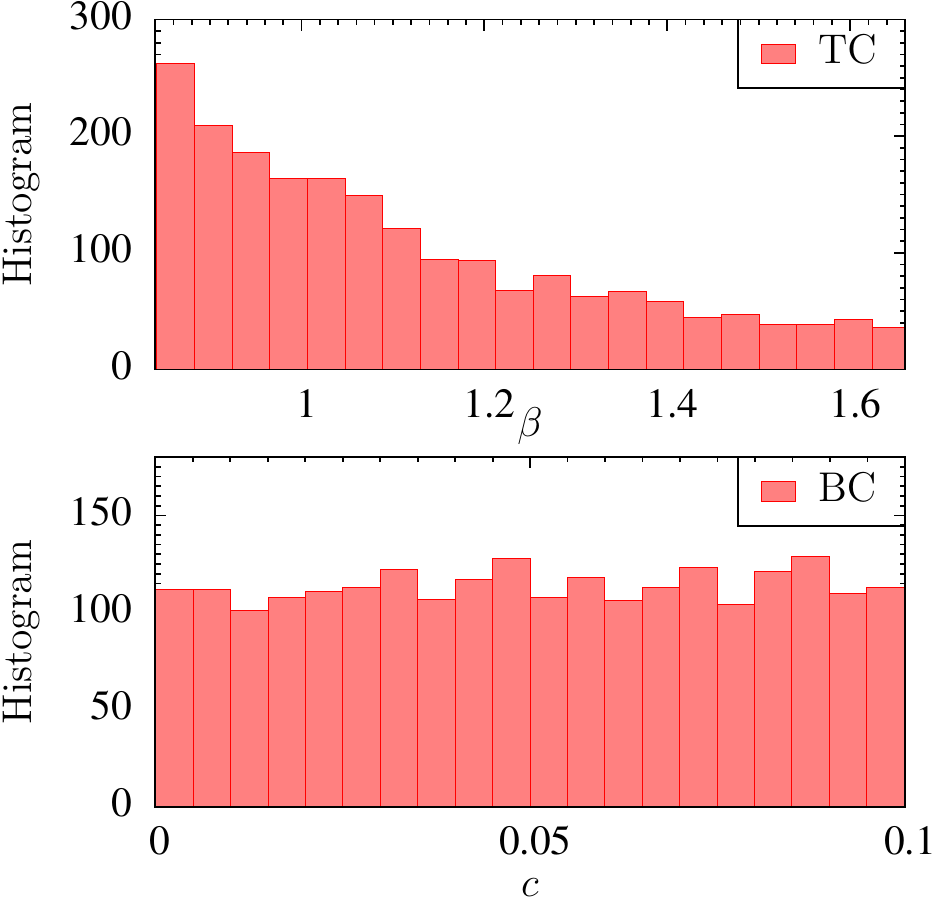}
\caption{
Distribution of all crossings of $L=6$ above a threshold $p_c$ for temperature chaos ($p_c=0.05$, upper panel) and bond chaos ($p_c=0.1$, lower panel), respectively. Temperature chaos is approximately exponentially distributed, and bond chaos is uniformly distributed. These distributions are similar to other sizes and also the three dimensions. %Bond chaos is much strong than temperature chaos, note the different horizontal scales.
}
\label{Ncrossing}
\end{center}
\end{figure}

Chaos in (partial) thermal boundary conditions manifests as crossings of boundary condition weights, as shown in Fig.~\ref{BCC} for two typical moderately chaotic instances of size $L=6$. The red circles and blue squares are examples of dominant and not dominant crossings, respectively.
%In our data analysis, we only use data when a crossing is above a chosen threshold $p_c$ for better accuracy. For temperature chaos we use $p_c=0.05$ and for bond chaos we use $p_c=0.1$ because bond chaos are much stronger than temperature chaos. Note this difference will not affect the number of dominant crossings, as no dominant crossing can occur below $p_c$ in both cases.
The histograms of those crossings above $p_c$ for all instances of $L=6$ are shown in Fig.~\ref{Ncrossing}. The distribution is approximately exponential with respect to $\beta$ for temperature chaos, while uniform with respect to $c$ for bond chaos. Our results clearly show that the effectiveness of temperature chaos \textit{decreases} rapidly with decreasing temperature in the spin-glass phase. 
%The distribution presumably indicates that the number of pairs of pure states that are separated by $\Delta F \sim O(1)$ (available for temperature chaos) decreases approximately exponentially with $\beta$ for a fixed $N$. 
The uniform distribution of bond chaos is easy to understand because of the statistical symmetry of $c$. The distributions are also very similar to their 3D counterparts. See the next section for more quantitative comparisons.

\begin{figure}[htb]
\begin{center}
\includegraphics[width=\columnwidth]{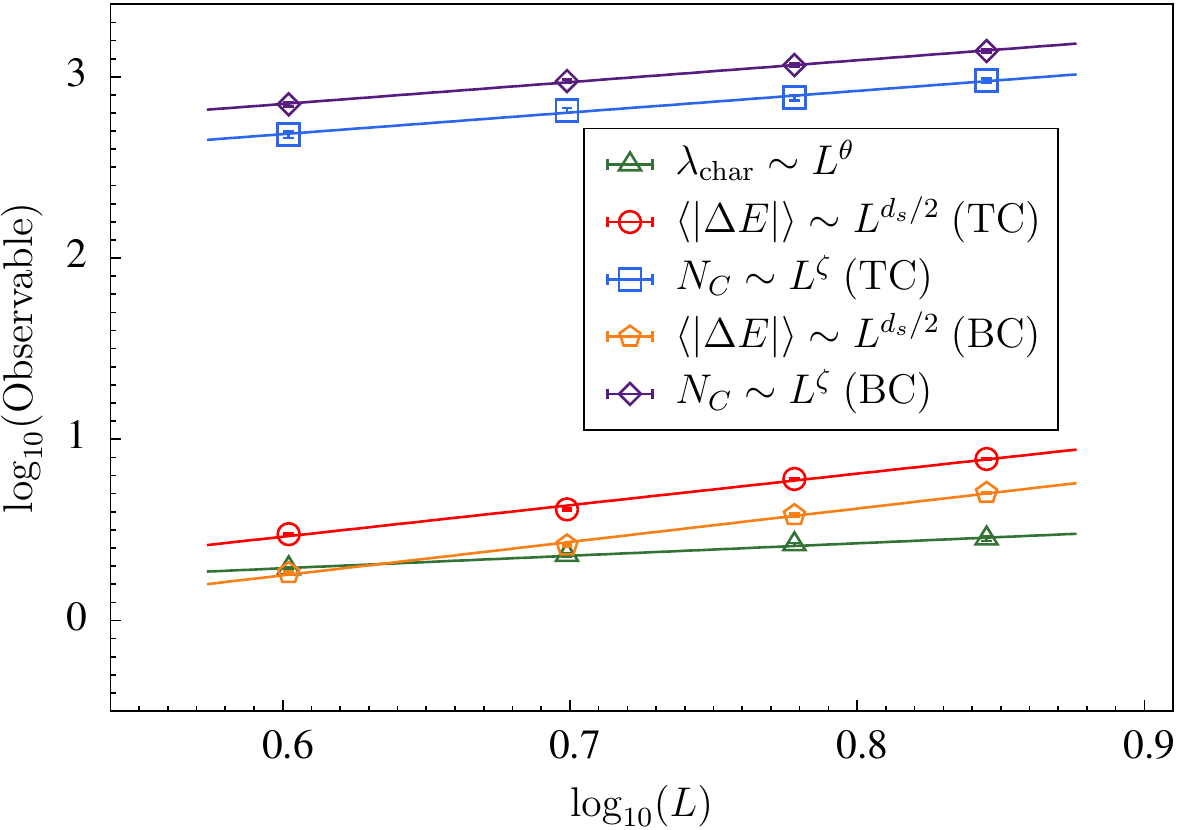}
\caption{
Scaling of the measured quantities as a function of the system size. The
log-log plot clearly shows that the different quantities are well fitted
with a power law. Sample stiffness scales as $\lambda_{\rm char} \sim L^\theta$, the energy difference at all boundary
condition crossings scales as $\Delta E \sim L^{\ds}$, and the number of dominant crossings scales
as $N_C \sim L^\zeta$. Error bars are smaller than the symbols.}
\label{SC2}
\end{center}
\end{figure}

One of our main results, the scalings of the sample stiffness $\lambda_{\rm{char}}$, $\langle |\Delta E| \rangle$ at crossings, and the total number of dominant crossings $N_C$ are shown in Fig.~\ref{SC2}.
Here, we have combined data at $(\beta=2/T_C,c=0)$ and $(\beta=2/T_C,c=0.1)$ to compute $\theta$ to
improve statistics, as the data at different $c$ are statistically equivalent.
Our estimates of the exponents are:
\begin{eqnarray}
\theta &=& 0.69(6) 				\\
\ds    &=& 1.74(3) \;\;\;\;\;\; {\rm (TC)}   	\\
\ds -\theta    &=& 1.05(7) \;\;\;\;\;\; {\rm (TC)}   	\\
\zeta  &=& 1.19(7) \;\;\;\;\;\; {\rm (TC)}   	\\
\ds    &=& 1.84(4) \;\;\;\;\;\; {\rm (BC)} 	\\
\ds -\theta    &=& 1.15(7) \;\;\;\;\;\; {\rm (BC)}   	\\
\zeta  &=& 1.20(6) \;\;\;\;\;\; {\rm (BC)}.
\end{eqnarray}

The agreement of the exponents for TC and BC are reasonably good, and both are compatible with the relation $\zeta = \ds -\theta$. Therefore, we conclude temperature chaos and bond chaos share the same set of scaling exponents in four dimensions, as in three dimensions \cite{Wang:BC}. The results also at the same time validate the partial thermal boundary condition technique for studying chaos.

Our estimate $\ds$ for TC is, however, somewhat smaller than that of BC, while the agreement of $\zeta$ is excellent. One possible reason for this result is that there might be larger systematic errors for temperature chaos when averaging $\langle |\Delta E| \rangle$ over a wide temperature range. In bond chaos, all quantities are averaged at a \textit{single} temperature. 
%From Fig.~\ref{Ncrossing}, we see that the BC data is measured at $\beta=2/T_C \approx 1.1$, which is to the high temperature side of the TC range. 
By narrowing down the temperature chaos range at low temperatures to only $T=0.8$ or $\beta=1.25$, the TC data set gives $\ds=1.77(4)$, in good agreement with the BC result. Therefore, we believe our BC estimate of $d_s$ is cleaner, and hence also the checking of the chaos equality. It is indeed the case that the relation $\zeta = \ds -\theta$ is in better agreement for bond chaos.

We now compare our results with the literature. Our stiffness exponent $\theta=0.69(6)$ is in good agreement with 0.61(2) using percolation method \cite{Boettcher:theta} and 0.64(5) using approximate ground states \cite{hartmann:99a}, both working at $T=0$ for the $\pm J$ model. Our estimate $d_s$ is also in agreement with a recent result $d_s \approx 3.74$ using a strong disorder renormalization group method \cite{Wang:fractal2}.
The chaos results are similar to that of Ref.~\cite{Hukushima:Chaos}, where chaos is studied for the $\pm J$ model using correlation functions: $\theta=0.69(3)$, $\ds=1.71(3)$, $\zeta=1.12(5)$ for temperature chaos and $1.10(10)$ for bond chaos. Our chaos exponents are slightly larger, but within errorbars.

\begin{table*}
\caption{
Summary of exponents (relevant to chaos) of the four-dimensional EA model. Note that not all of these work are for studying chaos, but a few typical related results are presented for comparisons of $\theta$, $d_s$ and $\zeta$. Here, MC and GS stand for ``Monte Carlo'' and ``Ground state'', respectively. We conclude that temperature chaos and bond chaos share the same set of chaos exponents, and the 4D EA spin glasses of Gaussian and $\pm J$ disorder also share the same set of chaos exponents.
\label{table}
}
\begin{tabular*}{\textwidth}{@{\extracolsep{\fill}} l c c r}
\hline
\hline
$\rm{Reference}$ &$\rm{model}$ &$\rm{result}$  &$\rm{note}$ \\
\hline
Ref.~\cite{neynifle:98} &Gaussian &$\zeta=0.85(10)\ (\rm{TC}),\ \zeta=0.95(20)\ (\rm{BC})$  &MC, $T=T_C=1.8$ \\
Ref.~\cite{neynifle:98} &Gaussian &$\zeta=1.2(1)\ (\rm{BC})$  &MC, $T=1.4$ \\
This work &Gaussian &$\zeta=1.19(7)\ (\rm{TC}),\ \zeta=1.20(6)\ (\rm{BC})$  &MC, $T \approx T_C/2 = 0.9$ \\
This work &Gaussian &$\theta=0.69(6),\ d_s/2=1.74(3)\ (\rm{TC}),\ d_s/2=1.84(4)\ (\rm{BC})$  &MC, $T \approx T_C/2 = 0.9$ \\
Ref.~\cite{Hukushima:Chaos} &$\pm J$ &$\zeta=1.12(5)\ (\rm{TC}),\ \zeta=1.10(10)\ (\rm{BC})$  &MC, $T \approx 0.6$ \\
Ref.~\cite{Hukushima:Chaos} &$\pm J$ &$\theta=0.69(3),\ d_s/2=1.71(3)\ (\rm{TC})$  &MC, $T \approx 0.6$ \\
Ref.~\cite{Boettcher:theta} &$\pm J$ &$\theta=0.61(2)$  &percolation \\
Ref.~\cite{hartmann:99a} &$\pm J$ &$\theta=0.64(5)$  &approximate GS \\
Ref.~\cite{Wang:fractal2} &Gaussian &$d_s=3.7358(36)$  &approximate GS \\
\hline
\hline
\end{tabular*}
\end{table*}

One earlier work with Gaussian disorder is Ref.~\cite{neynifle:98}. The author, however, separated two cases: Chaos at $T_C$ and Chaos below $T_C$. The results are $\zeta=0.85(10)$ for temperature chaos and $\zeta=0.95(20)$ for bond chaos at $T_C$. Notice that they are compatible, even though they may differ from the exponent in the spin-glass phase. Below $T_C$ in the spin-glass phase, only bond chaos was studied and $\zeta=1.2(1)$ at $T=1.4$, also for sizes up to $L=7$. This exponent is remarkably in good agreement with our results, even though the temperature is higher. All of these exponents are summarized for convenience in Table.~\ref{table}.
Taking all these results collectively, we conclude also that the $\pm J$ model has the same scaling exponents with Gaussian disorders in four dimensions.

\subsection{Relative strength of chaos}
\label{compare}

Next, we compare the relative strength of temperature chaos and bond
chaos at $T_C/2$, and also compare with that of 3D. We define density of crossings for both chaos, and the relative strength can be quantified as ratio of the densities. We follow the procedures established in Ref.~\cite{Wang:BC} for three dimensions.
The density of crossings for bond chaos is given by
\begin{equation}
\rho^{\rm BC}  = \frac{N_C}{\beta \Delta c}.
\label{eq:aa} 
\end{equation}
The distribution for temperature chaos is more complicated and is
approximately exponential in the range $\beta \in [\beta_{\rm
min},\beta_{\rm max}] = [3/(2T_C),3/T_C]$ for all system sizes studied. An exponential fit of the dominant crossing distributions of all sizes taken together of the form
\begin{equation}
f(\beta) = 
\frac{a e^{-a \beta}}{e^{-\beta_{\rm min} a} - e^{-\beta_{\rm max} a}} 
\end{equation}
in the temperature range yields $a \approx 1.78$. This exponent is appreciably larger than that of 3D $a \approx 1.12$ in the same relative range $\beta \in [3/(2T_C),3/T_C]$.
With this density distribution, we can easily compute the density at $\beta=2/T_C$ is approximately $1.17$ times of the average density in the full temperature range.
The corresponding density of crossings for temperature chaos at $T_C/2$ is therefore given by
\begin{equation}
\rho^{\rm TC} = \frac{1.17N_C}{\Delta \beta}.
\label{eq:bb}
\end{equation}
Remarkably, the prefactor 1.17 depends only very weakly on $a$ and is very similar to that of the 3D 1.18 where bond chaos is again also studied at $\beta=2/T_C$ \cite{Wang:TC,Wang:BC}. This therefore, provides also an excellent setting to compare the relative strength with the three dimension, as we will do in the following.

\begin{figure}[htb]
\begin{center}
\includegraphics[width=\columnwidth]{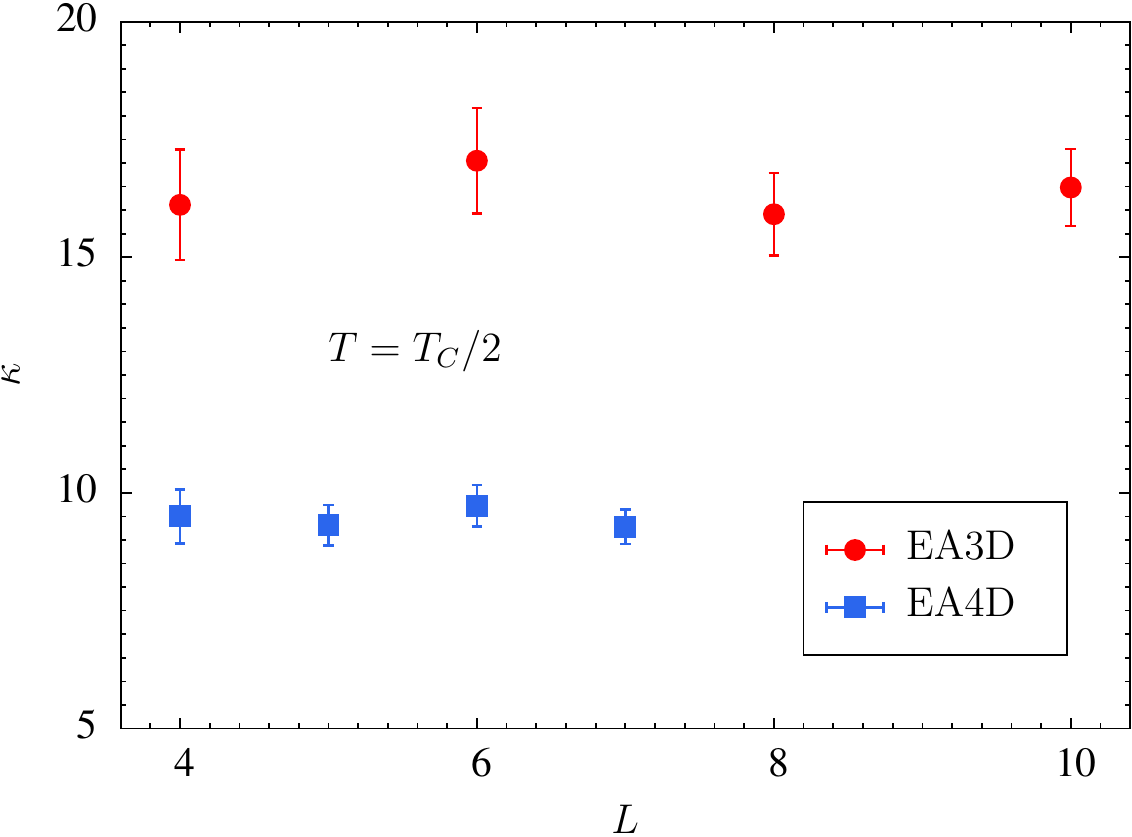}
\caption{
Relative strength between bond and temperature chaos $\kappa$ at $T_C/2$, as a
function of the system size $L$ for both three dimensions (red circles) and four dimensions (blue squares). Note that $\kappa$ is approximately size independent as expected from the scaling relations. The relative strength at the same relative temperature $T_C/2$ is, however, not a universal constant for different dimensions. Nevertheless, bond chaos in both cases are typically one order of magnitude stronger than temperature chaos.
}
\label{kappa}
\end{center}
\end{figure}

The relative strength of bond chaos to temperature chaos is naturally defined as:
\begin{eqnarray}
\kappa &=&       \frac{\rho^{\rm BC}}{\rho^{\rm TC}}, \nonumber \\
       &\approx& 6.41 \frac{{N_C}^{\rm BC}}{{N_C}^{\rm TC}},
\label{E1}
\end{eqnarray}
where ${N_C}^{\rm BC}$ and ${N_C}^{\rm TC}$ are the total number of
dominant boundary condition crossings of bond chaos and temperature
chaos, respectively. The prefactor is again similar to three dimensions, where it is 6.34 \cite{Wang:BC}. A plot of $\kappa$ as a function of the linear
system size $L$ is shown in Fig.~\ref{kappa}.

Firstly, $\kappa$ is almost a constant function of $L$, as expected from the scaling properties of $N_C$. The interesting finding is that the relative strength is not a universal constant at the same scaled temperature $T=T_C/2$. Averaging over all studied system sizes, we get $\kappa=9.5(1)$, compared with that of three dimensions $16(1)$. Ref.~\citep{Hukushima:Chaos} got a value $17.5$ for the 4D $\pm J$ model. While this appears to be rather close to the value obtained in three dimensions instead as observed in Ref.~\cite{Wang:BC}, this is likely an interesting coincidence rather than suggesting the ratio is a universal constant at a typical low temperature. This large value is not in disagreement with our data as the value is calculated at a relatively lower temperature $0.3T_C$. It is expected that $\kappa$ should increase with $\beta$. For example, bond chaos should persist even at $T=0$, while for temperature chaos, this is likely negligibly small for the finite sizes we have studied. Nevertheless, all of these data are fairly close, suggesting that bond chaos at a \textit{typical} low temperature is almost an order of magnitude stronger than temperature chaos.

It is possible to qualitatively explain why the ratio $\kappa$ is smaller for the 4D case than that of 3D in our studies where both are at $T_C/2$. It is presumably a result of the increased entropy relative to the energy in 4D. Here, we have looked at two quantities and find this appears to be the case. Our first quantity is based on the overlap distribution function $P(q)$ where the overlap $q$ in our thermal ensemble is defined as
\begin{equation}
q_{ab} = \dfrac{1}{N} \sum_i S_i^a S_i^b,
\end{equation}
where the two replicas $a, b$ are chosen randomly (including the boundary conditions) from the TBC ensemble. The overlap distribution function quantifies the similarities of the different states, or pure states in the thermodynamic limit. The overlap distribution is trivial if there is only one pair of pure states and is nontrivial when there are many pairs of pure states. We compute an extensively used statistic which is the cumulative integral of the function near $q=0$ as
\begin{eqnarray}
I(0.2) = \int_{-0.2}^{0.2} P(q) dq.
\end{eqnarray}
The disorder average is well-known to be approximately a constant function of $L$. This provides a definition of the \textit{effective} relative temperature \cite{billoire:13} of the system again with respect to $T_C$ based on the strength of excitations in the spin-glass phase. The statistic $I(0.2)$ equals $0.1302(54)$ and $0.1805(40)$ in 3D and 4D, respectively. Therefore, the 4D data is at a higher effective relative temperature than the 3D data, which explains why the 4D $\kappa$ is smaller. The ratio of the two is $1.39(7)$ which is approximately of the same scale as the ratio of $\kappa$ which is $1.68(11)$. The other quantity we looked at is the direct ratio of the energy to entropy scales $[\langle E \rangle ] / [\langle TS \rangle ]$ at $T_C/2$, where the square brakets denote disorder averages. The entropy is computed from the energy and the free energy which can be easily measured in population annealing using the free energy perturbation method \cite{Wang:PA}. The estimates are $56(3)$ and $22.0(3)$ for 3D and 4D, respectively.
%Note that the fluctuations from size to size is a lot smaller in 4D.
The ratio of the two is $2.54(15)$ which is again approximately of the same scale as the ratio of $\kappa$. It is important to emphasize, however, that both quantities are merely estimates of scales. Neither is expected to be an estimator of $\kappa$. Nevertheless, it appears relatively clear and we conclude that $\kappa$ gets smaller in 4D than 3D at $T_C/2$ as a consequence of the increased entropy relative to energy.

%We propose a simple physical interpretation for $\kappa$. At first
%sight, one might expect that the scale of $\Delta E$ and $\Delta(TS)$
%might be relevant to explain $\kappa$. However, while we find this is
%indeed a factor---especially at low temperatures---this is not
%sufficient. We find that the strength of \textit{responses} of the
%quantities with respect to $c$ and $\beta$ are more relevant, as chaos
%is a dynamical processes. To this effect, we use an alternate definition
%of the relative strength $\kappa$, namely
%\begin{eqnarray}
%\kappa &=& \left\langle
%		\frac{\partial |\Delta E|}{\beta \partial c} 
%	   \right\rangle 
%      \Big{/} 
%	   \left\langle
%		\frac{\partial |\Delta(TS)|}{\partial \beta} 
%	   \right\rangle,
%\label{E2}
%\end{eqnarray}
%evaluated at the same temperature $\beta=2$. This alternate definition
%of $\kappa$ is also shown in Fig.~\ref{eta} (blue squares). Note that
%Eq.~\eqref{E2} also explains why $\kappa$ does not depend on the system
%size \cm{from} the \cm{same} scaling properties of $\Delta E$ and $\Delta(TS)$.
%The predictions are reasonably close, showing that bond chaos is indeed
%energy driven, while temperature chaos is entropy driven.

\subsection{Does chaos imply many pure states?}
In this section, we discuss whether chaos would imply a nontrivial overlap distribution in the framework of thermal boundary conditions. It may seem inconsistent that we have employed the droplet description of chaos and now argue against it. However, we are here only questioning the number of pure states, not its scaling description of chaos. Indeed, we argue in the following that many states and the droplet description of chaos can also be consistent.
%it is possible to modify the droplet picture to many states without affecting its applicability to chaos. %which in turn could be a stepping stone towards formulating a complete picture that is fully consistent with all simulations for the EA model.

Firstly, the droplet scaling of chaos is scaling with respect to the system size $L$, and does not require that there are only two pure states. Similar to our finding that the number of boundary conditions does not affect the scaling exponents, we expect the same is true for pure states as well only provided that the effective number of active pure states (not with a vanishingly small weight) should be about the same for different $L$. Recall that chaos refers to or is dominated by large-scale reorganizations. This is indeed the case in a many-state picture because despite there are many (a countable infinity in the thermodynamic limit) pure states, only a handful of them have $O(1)$ weights \cite{book}. This is also reflected in that the pool of the overlap distribution functions $\{P_{J}(q)\}$ looks similar for different sizes like the aforementioned statistic $I(0.2)$. Therefore, there is no apparent inconsistency between many pure states and the validity of the droplet description of chaos. The droplet description of chaos could be applied to any pair of those active pure state exchanges. Finally, many pure states would, while not affecting the three scaling exponents, clearly enhance the intensity of chaos or the prefactor of this scaling.

Next, we discuss why we consider the possibility of many pure states. The droplet picture \cite{fisher:86,fisher:87,fisher:88,bray:86,mcmillan:84b} has long been believed to be a two-state picture, as the exponent $\theta_{\rm{DW}}>0$ assuming droplet excitations and domain-wall excitations are similar in nature. However, numerical simulations have been observing nontrivial overlap distributions, i.e., many pure states. There is so far no direct evidence that the overlap distributions are trivial. This is either interpreted as evidence for the replica symmetry breaking (RSB) picture \cite{parisi:79,parisi:80,parisi:83} or as a finite-size effect. 
%It is beyond the scope of this paper to discuss fully this controversial topic, but now we focus on this question in the framework of thermal boundary conditions. %However, we believe the boundary condition crossing characterization of chaos may be very useful for answering this question and therefore
It seems more likely the former is correct, as it is actually questionable that $\theta_{\rm{DW}} > 0$ would imply absence of large-scaling excitations for \textit{all} instances. For example, even the mean-field Sherrington-Kirkpatrick model \cite{sherrington:75} appears to have a positive exponent $\theta_{\rm{DW}}$, but the model is clearly described by RSB \cite{Wang:KAS}. In addition, $\theta_{\rm{DW}}$ appears to be simply a growing function of dimensionality and remains positive such as at $D=7$ \cite{boettcher:05d} which is already above the upper critical dimension presumably $D=6$. In the following, we discuss a \textit{tentative} view that the two-state picture may not hold from the perspective of chaos in the TBC ensemble. We propose a picture that results in both a positive exponent $\theta_{\rm{DW}}$ and nontrivial overlap distributions.

In fact, the primary motivation of the TBC \cite{Wang:TBC} is exactly to address the number of pure states. Reference~\cite{Wang:TBC} did find nontrivial overlap distributions from \textit{direct} computations, but instead concluded the overlap distributions should become trivial using an \textit{indirect} sample stiffness scaling. The basic idea is that more stiff instances (large $\lambda$, one dominant boundary condition) are found to be correlated with more trivial overlap distributions (small $I(0.2)$) and all instances are argued to become infinitely stiff ($\lambda \rightarrow \infty$) in the thermodynamic limit, similar to the above mentioned droplet picture. The correlation looks robust, but the latter is questionable. The paper indeed stated that this may not occur if a finite fraction of instances get increasingly more stiff with $\Delta F \sim L^{\theta}$ while the others do not with $\Delta F \sim O(1)$. This scenario was simply rejected as there had been no straightforward explanation to expect this, but chaos appears to provide such a picture as we discuss below.

\begin{figure}[htb]
\begin{center}
\includegraphics[width=\columnwidth]{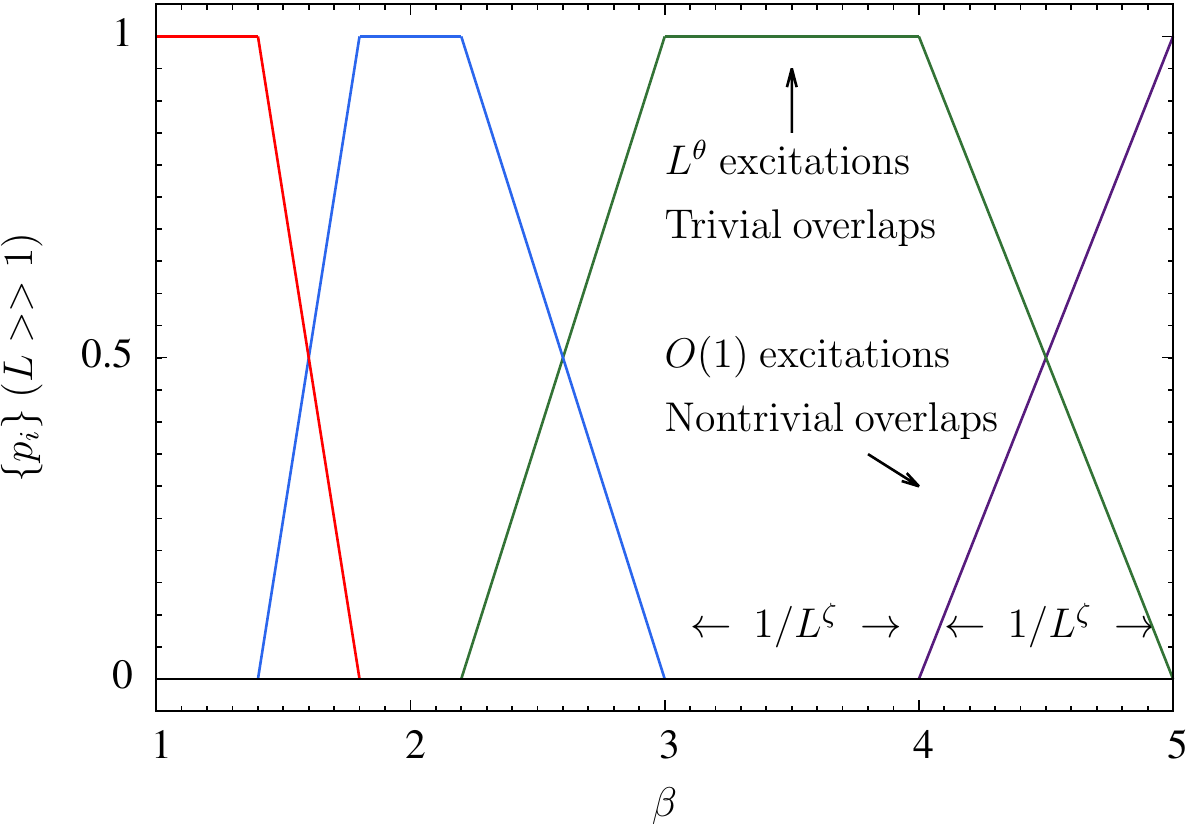}
\includegraphics[width=\columnwidth]{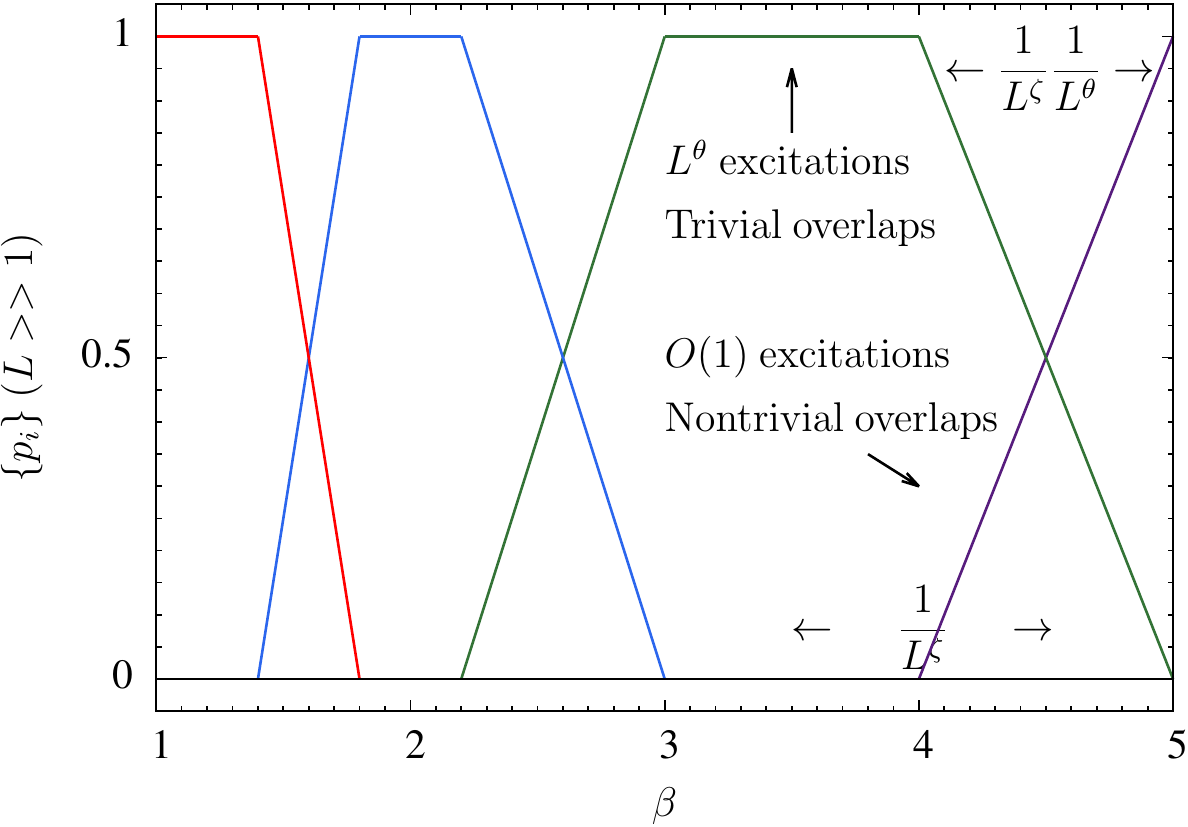}
\caption{
Top panel: One possible scenario of temperature chaos in the TBC ensemble in the thermodynamic limit.
% Note that the temperature range is partitioned into two different regimes. 
%If $N_C \sim L^{\zeta}$, each crossing event should scale as $1/\ell^{\zeta}$. 
Each exchange event has two regimes: an $O(L^{\theta})$ regime and an $O(1)$ regime in terms of free energy differences. In the former regime, one boundary condition dominates and the overlap distribution is trivial (without considering excitations within the same boundary conditions). In the latter regime, two boundary conditions have comparable weights and the overlap distribution is nontrivial. The width of both regimes scales as $1/L^{\zeta}$. The picture has a positive exponent $\theta$ but a nontrivial overlap distribution function after taking disorder averages at any arbitrarily chosen temperature, in agreement with numerical simulations. Bottom panel: Another possible scenario to save the droplet picture as a two-state picture \cite{comment:machta}. In this picture, the two regions share very unequal weights, the crossing region shrinks by an additional factor $1/L^{\theta}$ compared with the flat region. In this way, the scaling relation $I(0.2) \sim 1/L^{\theta}$ is recovered.
}
\label{BCC4}
\end{center}
\end{figure}

Our consideration is motivated by the following question: Suppose in the thermodynamic limit, one boundary condition dominates the ensemble as required by the droplet picture, but not the same one as temperature varies. If the boundary conditions are constantly exchanging their dominance, why would we always see one boundary condition whenever we measure their weights? We therefore propose the following picture for the thermodynamic limit as shown in the top panel of Fig.~\ref{BCC4}. Clearly if $N_C \sim L^{\zeta}$, each exchange event defined from a central maximum to a nearby central crossing should scale as $1/L^{\zeta}$. Each exchange event has two regimes: an $O(L^{\theta})$ regime and an $O(1)$ regime in terms of free energy differences. In the former regime, one boundary condition dominates and the overlap distribution is trivial. In the latter regime, two (or more perhaps with a smaller probability) boundary conditions have comparable weights and the overlap distribution is nontrivial. Motivated by the droplet scaling, we further propose the most natural scenario that the two regimes are of similar width and therefore they both scale as $1/L^{\zeta}$. Notice that in our analysis excitations within a single boundary condition are not considered, which would only make the overlap distributions even less trivial.

The advantage of this picture is that it is in agreement with all the aforementioned numerical results.
At an arbitrarily fixed temperature, an instance may be randomly observed in either regime. When taking disorder average, the exponent $\theta$ would be dominated by the $O(L^{\theta})$ regimes and on the other hand the overlap distribution function is dominated by the $O(1)$ regimes. This picture is also compatible with the distributions of $\lambda$ of Ref.~\cite{Wang:TBC} where the distribution is found to only change significantly at the tail of the distribution where $\lambda$ is large and the distribution at small $\lambda$ hardly changes. Therefore, our picture naturally provides a scenario of two different classes of instances, and a finite fraction of instances would not become stiff even in the thermodynamic limit.

The validity of this scenario depends crucially on the about equal share of the two regimes. We have recently indeed heard a possible way to save the droplet picture \cite{comment:machta} and it is shown in the bottom panel of Fig.~\ref{BCC4}. In this alternative picture, the $O(L^{\theta})$ regime in each exchange event takes most of the share and the $O(1)$ regime has only a tiny share of $1/L^{\theta}$ (of the width $1/L^{\zeta}$), then the total length of the $O(1)$ regimes would shrink as $1/L^{\theta}$ and the droplet behaviour such as $I(0.2) \sim 1/L^{\theta}$ is recovered.
While this exotic scenario would again yield a two-state picture, we do not readily see an obvious reason for such uneven shares. For example, the inversion from Eq.~\ref{DQ} to Eq.~\ref{NC} would be much less straightforward in this scenario. Moreover, we do not seem to see such uneven shares and such a strong trend for the sizes we have studied. In the rest of this section, we use an effective statistic to quantitatively distinguish the two scenarios.

\begin{figure}[htb]
\begin{center}
\includegraphics[width=\columnwidth]{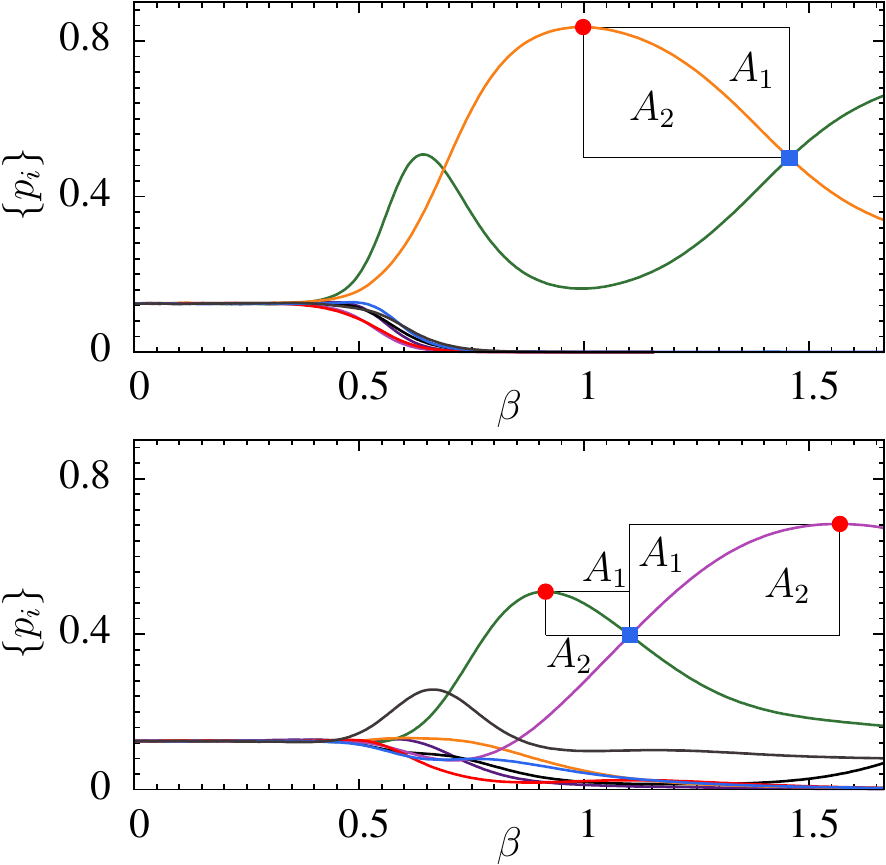}
\caption{
A dominant maximum (red circles) and a dominant crossing (blue square) constitute a dominant exchange event (black box), and examples are typical instances with such dominant exchanges chosen from $L=7$. If the maximum occurs at a smaller $\beta$, it is defined as a forward exchange. On the other hand, it is defined as a backward exchange. We define a statistic $\gamma = A_1/(A_1+A_2)$ to distinguish the two scenarios shown in Fig.~\ref{BCC4}.
}
\label{BCC6}
\end{center}
\end{figure}

It is clear that our sizes are far away from the limit where only two boundary conditions dominant, therefore it is of crucial importance to design a good statistic that is not very sensitive to this to look for a trend. Since we are basically interested in the shape of the curves, we define a statistic $\gamma$ to quantify the shape or the concavity of the probability curves of such exchange events. Firstly, we define a dominant exchange event. We have already defined a dominant crossing, now we define a dominant maximum which is a local maximum of a dominant boundary condition. We define a dominant exchange as such a maximum and its nearest dominant crossing. Some typical examples of these are shown as block boxes, red circles and blue squares, respectively in Fig.\ref{BCC6}. Such exchanges are the finite versions of the exchanges shown in Fig.~\ref{BCC4}.
We numerically integrate the area below the probability curve in the box $A_2$. The area above the curve $A_1$ can also be easily computed as the total area $A=A_1+A_2$ can be easily computed. We define
\begin{eqnarray}
\gamma = \dfrac{A_1}{A_1+A_2},
\end{eqnarray}
which captures the relative width of the two regimes or the sharpness of the crossings shown in Fig.~\ref{BCC4}. More precisely, we expect
\begin{eqnarray}
\gamma &=& \rm{const} \in [0,0.5] \rm{\hspace{1cm} (equal\ shares)}, \\
\gamma &\sim& 1/L^{\theta} \rm{\hspace{2.1cm} (unequal\ shares)}.
\end{eqnarray} 
In practice, we study the exchange events in the interval $\beta \in [3/(2T_C),3/T_C]$, and we require also the size of an exchange event to satisfy $\Delta \beta \geq 0.1$ and $\Delta p \geq 0.02$ for the purpose of numerical accuracy.

\begin{figure}[htb]
\begin{center}
\includegraphics[width=\columnwidth]{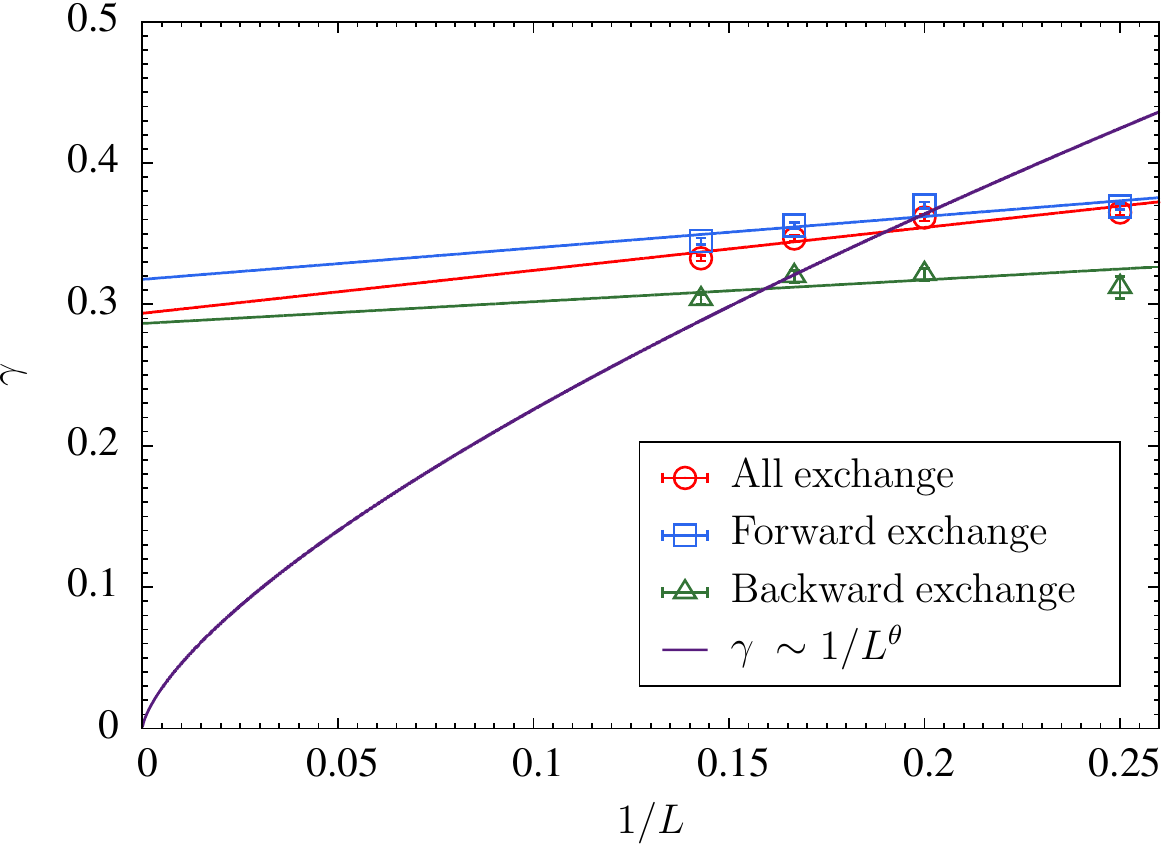}
\caption{
The statistic $\gamma$ as a function of $1/L$ along with linear fits and the droplet fit. The small decrease of $\gamma$ with increasing $L$ is most likely finite-size effect from the subdominant boundary conditions. The droplet fit does not appear to fit the data very well. See the text for more details.
}
\label{gamma}
\end{center}
\end{figure}

The results of $\gamma$ as a function of $1/L$ are shown in Fig.~\ref{gamma} along with a linear fit and the droplet fit. The values of $\gamma$ are converging to a constant that significantly differs from $0$ for the leading linear fit. The droplet fit which requires $\gamma \rightarrow 0$ in the thermodynamic limit, on the other hand, gives a poor fit, questioning the rapid shrinking of the crossing regions in the second scenario of Fig.~\ref{BCC4}. In the droplet fit, we have restricted the value of $\theta$ to our earlier estimate.

It does appear $\gamma$ is decreasing slightly as $L$ is increased, although very slowly and also by limited amounts. We attribute this to finite-size effects of the subdominant boundary conditions. To illustrate this, we divide the exchanges to two classes: Forward exchanges and backward exchanges. If the maximum occurs at a smaller $\beta$ or the dominant boundary condition is losing weight, it is a forward exchange. On the other hand, if the maximum occurs at a larger $\beta$ or the dominant boundary condition is gaining weight, it is a backward exchange. Note that due to the ``boundary condition'' $p_i = 1/8$ at $\beta=0$, it is most likely to encounter a forward exchange first than a backward exchange when $\beta$ is increased. The reason we do this classification is because the effects of the subdominant boundary conditions on $\gamma$ are opposite in these two cases. Consider the forward exchange, the dominant boundary condition loses weight, it is statistically more likely the subdominant ones (the little ones at the bottom of the probability curves like the forests in the bottom panel of Fig.~\ref{BCC6}.) are gaining weights. This would make the dominant boundary condition have a less concave shape and as a result $\gamma$ gets larger. Similar arguments show that $\gamma$ tends to be smaller for the backward exchange. One finite-size effect comes in when considering that a larger size is more likely to produce a backward exchange, because it is more chaotic. It is less likely for $L=4$ to make a backward exchange following a forward exchange, but $L=7$ can make this more frequently. We have looked at the fractions of such backward exchanges, which is indeed an increasing function of $L$. The fractions are $0.0743, 0.1794, 0.2574, 0.2949$ for $L=4, 5, 6, 7$, respectively. The averages using only forward exchanges or backward exchanges are also shown in Fig.~\ref{gamma}. It is clear that the forward exchanges are larger and backward exchanges are smaller, in agreement with our expectations. Ideally, these two averages should be flat now, both are certainly more flat than the full average. However, there is an additional finite-size effect that the subdominant boundary conditions are getting suppressed as $L$ increases, which is why we count only dominant crossings in our study of chaos. This is more pronounced for the forward class because they tend to occur at higher temperatures, which explains why smaller sizes deviate further from the thermodynamic limit in the forward class. On the other hand, the backward class is more flat because they tend to occur at lower temperatures, where the effects of subdominant boundary conditions should be smaller. Therefore, we believe the subdomiant boundary conditions are the source of the finite-size effects and the backward average is closer to the thermodynamic limit. In conclusion, our data of $\gamma$ is more consistent with many pure states ($\gamma=$const) with minor finite-size corrections from the subdominant boundary conditions, and does not fit the droplet two-state picture ($\gamma \sim 1/L^{\theta}$) very well.

%Perhaps a better way to address this problem is to study
%But the picture of two classes of free energy differences and a trivial or nontrivial overlap distribution function can be tested in the future with the simper case of using PBC and APBC only in the $x$-direction, as free energy is now routinely measured in spin-glass simulations \cite{Machta:PA,Wang:PA}.

%we may randomly see an instance with or without large-scale excitations. We suspect this is how an instance generate complex overlap distributions. In numerical simulations, we do see some instances having overlap functions that are very clean, while others with some extra peaks besides the Edwards-Anderson order parameter. For the domain-wall free energy exponent, this would naturally give $\theta > 0$ as the disorder average is dominated by costly domain walls, not the $O(1)$ ones.

%In summary, it appears we can state the following:
%\begin{enumerate}
%\item Chaos modifies the droplet picture to have complex overlap distributions.
%\item Domain walls and droplets are similar in free energy cost and both have two types: costly ones and soft ones. In particular, domain walls and droplets of $O(1)$ free energy costs exist with nontrivial probabilities. Note that there are obviously costly droplets too, though they are less interesting perhaps. If they also have the same geometry, then domain walls and droplets have exactly the same properties. The costly ones scale as $L^{\theta}$, which dominant the scaling.
%\end{enumerate}

\section{Conclusions \& future challenges}
\label{cc}

In this work, we have successfully extended the thermal boundary condition technique to partial thermal boundary conditions, and applied it to study the temperature chaos and bond chaos of the four-dimensional Edwards-Anderson model with Gaussian disorder to low temperatures. We have measured the three scaling exponents of chaos, and found with good accuracy that they are related through the chaos equality of the droplet picture and the two forms of chaos share the same set of scaling exponents. Our results and the literature values also suggest that the scaling exponents are the same for the Gaussian disorder and the $\pm J$ model in four dimensions, unlike two dimensions. Quantitative comparison of the relative strength of bond chaos and temperature chaos are also made at $T=T_C/2$ and compared with 3D. The relative strength is found to be slightly smaller but still similar in 4D and this is explained as the increase of entropy relative to energy in 4D. Temperature chaos distributions in 3D and 4D are also qualitatively similar, but nonetheless also quantitatively different, where 4D has a larger exponent in the exponential distribution. Finally, we have proposed a tentative scenario that chaos may imply many pure states in the TBC ensemble. This picture agrees with the numerical results of the TBC ensemble, and it is consistent with the scaling properties of chaos, a positive domain-wall exponent and also many pure states.

Our results pave the way for the (partial) thermal boundary condition technique to be applied to a wide range of models, as the number of fluctuating boundary conditions can be chosen flexibly (up to factors of 2). For example, it is possible to use the method efficiently to study chaos of one-dimensional long-range models on a ring such as the mean-field Sherrington-Kirkpatrick model \cite{sherrington:75} by also keeping 8 boundary conditions by introducing three equally-spaced points as boundaries. In particular, the model also has a spin-glass phase in a magnetic field, and therefore temperature chaos, bond chaos and field chaos can be characterized and compared on the same footing. It is also straightforward and interesting to apply the method to other spin-lattice models such as Potts, clock, XY and Heisenberg spin glasses. Chaos of these models are far less studied but may exhibit new interesting phenomena.
For example, the clock spin glasses can have an extremely rich phase diagram such as a chiral spin-glass phase, which is also chaotic \cite{Berker:clockglass}.
Finally, we look forward to seeing Monte Carlo simulations of the Edwards-Anderson model in yet higher dimensions as a result of Moore's law and parallel computing. Using the strong-disorder renormalization group $d_s$ \cite{Wang:fractal2} and the domain-wall stiffness exponent $\theta$ \cite{Boettcher:theta} and assuming the droplet description of chaos is correct up to 6D \cite{Wang:fractal,Moore:AT}, we estimate $\zeta=1.56(5)$ and $1.89(10)$ in five and six dimensions, respectively.

\acknowledgments
We thank J. Machta and M. A. Moore for helpful discussions.
W.W.~acknowledges support from the Swedish Research Council Grant No.~642-2013-7837 and Goran Gustafsson Foundation for Research in
Natural Sciences and Medicine. M.W.~acknowledges support from the Swedish Research Council Grant No.~621-2012-3984.
%Computations were performed on resources provided by the Swedish National Infrastructure for Computing (SNIC) on the Triolith cluster at NSC and the Kebnekaise cluster at HPC2N.
The computations were performed on resources
provided by the Swedish National Infrastructure for Computing (SNIC)
at the National Supercomputer Centre (NSC) and the High Performance Computing Center North (HPC2N).

%The computations/simulations/[SIMILAR] were performed on resources
%provided by the Swedish National Infrastructure for Computing (SNIC)
%at [CENTERNAME (CENTER-ACRONYM)].
%We thank Swedish National Infrastructure for Computing (SNIC) for access to the Triolith (NSC) and Kebnekaise (HPC2N) clusters.
%We thank Swedish National Infrastructure for Computing (SNIC) for access to their Triolith cluster. 
%(NSC) and Kebnekaise (HPC2N) clusters.

\bibliography{Refs}

\end{document}